\documentclass[conference]{IEEEtran}
\IEEEoverridecommandlockouts
\usepackage{cite}
\usepackage{amsmath,amssymb,amsfonts}
\usepackage{algorithmic}
\usepackage{graphicx}
\usepackage{textcomp}
\usepackage{xcolor}
\usepackage{color}
\usepackage{balance}
\def\BibTeX{{\rm B\kern-.05em{\sc i\kern-.025em b}\kern-.08em
    T\kern-.1667em\lower.7ex\hbox{E}\kern-.125emX}}

\begin{document}

\title{ \huge An Ultimate Approach of Mitigating Attacks in RPL Based Low Power Lossy Networks}

\author{\IEEEauthorblockN{Jaspreet Kaur }
\IEEEauthorblockA{\textit{PhD Scholar at CSE Department}
 \\
\textit{Indian Institute of Technology Jodhpur}\\
Jodhpur, India \\
kaur.3@iitj.ac.in}}
\maketitle
\begin{abstract}
The Routing Protocol for Low-Power and Lossy Networks
(RPL) is the existing routing protocol for Internet of Things
(IoT). RPL is a proactive,lightweight, Distance Vector protocol which
offers security against various forms of routing attacks. Still, there are various attacks(as rank, version attacks and many more ) which is possible in this network due to
problem of unauthenticated or unencrypted control frames,
centralized root controller, compromised or unauthenticated
devices and many more ways. There are various solutions
present in the literature but every solution has its pros and
cons. There is no appropriate system framework till now
which completely solves these all issues. So, we present an
ultimate approach to
mitigate these RPL attacks more efficiently and effectively. We use IDS based system for internal attacks and a mini-firewall for removing the external attacks. In IDS based approach, we use intrusion detection system at multiple locations for analyzing the behaviour of nodes. The final decision whether the node is attacker or not depends on mainly three  things as: trust between the neighbouring nodes, local decision by multiple sink nodes and global decision by root node. We also use some blockchain features in this framework for better internal security. We use some threshold values and rules  in mini-firewall for removing external attacks.  In
this paper, we provide the proposed approach and theoretical analysis of this approach
which provide better protection from these attacks
than any other method.
\end{abstract}

\begin{IEEEkeywords}
Internet of Things(IoT), Routing Protocol, RPL Attacks, Low Power Lossy Networks, Multiple Sink Nodes, Trust, Local Decision, Global Decision, Blockchain, Intrusion Detection System(IDS) , Mini-Firewall.
\end{IEEEkeywords}

\section{Introduction}

In today’s world, IoT is technical revolutionary area in
mobile and wireless communication field which deploy Low
power and Lossy Networks (LLN). These networks are typically
composed of many heterogeneous embedded devices
with limited power, memory, and processing resources. Now,
IoT is applicable in many areas such as industrial monitoring,
smart home, health care, environmental monitoring,
smart city, smart grid and many more. Due to the huge number of applications of these networks, security become a critical part for privacy of personal data. RPL is default routing protocol in these network which is susceptible from various attacks. Now, We briefly describe the related technologies use in this paper.

\subsection{Routing Protocol as RPL and it's Attacks}
RPL is distance based proactive protocol used for routing
in IoT network. At beginning, a RPL protocol creates a graph-like structure
called the Destination Oriented Directed Acyclic Graph
(DODAG). The DODAG consists of paths from the sender
nodes(IoT devices) to the sink node(6LBR). During routing, every node maintains its rank relative to its position in
the DODAG tree, and every DODAG is maintained by control
and route information . The control frames are used by
DODAG are DODAG Information Object (DIO),Destination
Advertisement Object (DAO) and DODAG Information Solicitation
(DIS) for transmitting the DODAG information.
Route path selection is a key factor for RPL, RPL use the
Various routing metrics, route constraints and objective functions
(OF) such as hop count, energy minimization and latency
to compute the best route path. The basic framework of RPL network is shown below.\\
\begin{figure}[h]
    \centering
    \includegraphics[width=10cm,height=9cm]{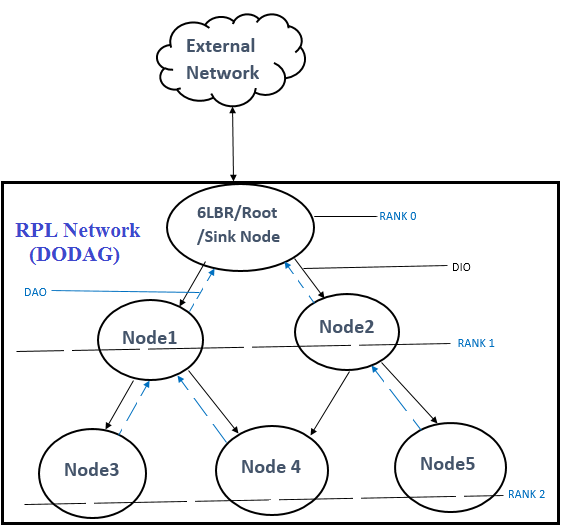}
    \caption{RPL Based Low Power Lossy Network}
    \label{fig:my_label}
\end{figure}

There are various attacks possible in RPL
network which significantly
impact the network resources and its performance.
 These attacks are possible due to the problem of
unauthenticated or unencrypted control frames, centralized
root controller, compromised and unauthenticated devices and many more ways. Some of these attacks are shown in figure 2 and briefly describe below.

\begin{figure}[h]
\centering
\includegraphics[width=9cm,height=9.5cm]{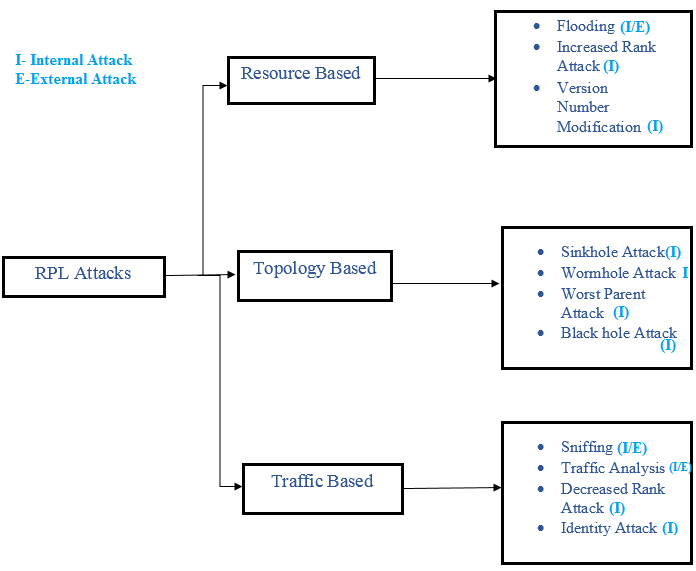}
    \caption{RPL Network Attacks}
    \label{fig:my_label}
\end{figure}

\subsubsection{Sinkhole Attack}
In this internal attack, attacker or compromised node advertises beneficial path to attract many nearby nodes to route traffic
through it. This attack disrupt the network topology and can become very stronger when combined with another
attacks[2].\\ 
\subsubsection{Version Number Modification Attack}
This internal attack is occurred  by changing version number(lower to higher) of a DODAG tree. When nodes receive the new
higher version number DIO message they start the formation of new DODAG tree. This results unoptimized or inconsistency of network
topology, increases control overhead and  higher packet loss[2].\\
\subsubsection{Denial of Service Attack}
Denial of service or Distributed denial of service attack is attempt to make resources unavailable to its
valuable  user [12-13]. In RPL this attack can be done by the IPv6 UDP packet flooding[2].\\
\subsubsection{Neighbor Attack}
In this attack, intruder broadcast DIO messages that it received without adding information of
himself.The node who receives this messages may conclude that new neighbor node send this DIO message. The victim
nodes select the out range neighbors node for routing purpose which affects network quality
parameters[2].\\  
\subsubsection{Wormhole Attack}
This attack can occurred  by creating tunnel between the two attackers and transmitting the selective traffic
through it which Disrupt the network topology and
traffic flow[2].\\
\subsubsection{Decreased Rank Attacks}
 In a DODAG, rank is used for positions of nodes relative to the sink node. It means lower the rank is close to the root and vice verse.  When a malicious node advertises a lower rank value, it means, it wants majority of traffic pass through it. As a result,
many legitimate nodes connect to the attacker. An attacker can change its rank value through the falsification of DIO messages.
finally, an attacker either drop this messages or selectively forward these messages. It becomes more powerful when combined with other attacks[2].\\ 
\subsubsection{Identity Attacks}
 Identity attacks contain clone ID and sybil attacks. In a clone ID attack, an attacker copies the identities of a
valid node onto another physical node.  In a sybil attack, which is similar to a clone ID attack, an attacker uses several logical entities
on the same physical node. These attacks can be used to take control over large parts of a network without deploying physical
nodes[2,12-13].\\
\subsubsection{Sniffing Attacks}
A sniffing attack perform by the listening of the packets transmitted over the network. This attack compromises the confidentiality of communications. It also check the pattern of traffic for major attacks[12-13].\\
\subsection{Blockchain}
Blockchain is fundamentally
a decentralized, distributed, shared, and immutable
database ledger that stores data across a peer-to-peer (P2P)
network. It has chained blocks of data that have been timestamped
and validated by miners. Fundamentally, the block
data contains a list of all transactions and a hash to the previous
block. The blockchain has a full history of all transactions
and provides a global distributed trust[1].
\subsection{Intrusion Detection System(IDS)}
An intrusion detection system (IDS) is a system that monitors network traffic for suspicious activity and issues alerts when such activity is occurred. These can be signature based or anomaly based and network or host based depending upon the need of application.  These mainly used for the detection of internal attacks.
\subsection{Firewall}
Firewall is used for protection from outsider attacks. This use some threshold values or specific rules to filter unwanted traffic such as filtering is based on port numbers, ip addresses and many more parameters. 
\subsection{Trust}
Trust means integrity, strength, ability, confidence of one person on the other person or thing. A trust is a relationship or agreement  which one party, known as a trustor, gives another party, the trustee.\\\\
The rest of the paper is organized as follows: section 2 describes
the Related Work and motivation for our work.
Section 3 mentions our proposed framework system followed
by the theoretical analysis of work in section 4. In Section 5, a conclusion of our findings is presented with future extensions
to our work.

\section{Related Work \& Motivation}
For detection and mitigation of RPL attacks, various
mechanisms are presented in the literature. But they
have their own pros and cons. They are useful for one attack but not for others. There are no standardized framework system for security of RPL network. Specific literature survey for some of the attacks are present below.\\\\
In paper[1], They describe various IoT security issues, open challenges and provide blockchain as a solution for these attacks. They also review the RPL attacks and their solutions present in the literature. In paper[2], the detailed survey of various RPL attacks and their solutions are presented. They describe taxonomy of  RPL attacks based on resources, topology and traffic. They also give risk management process for these attacks.\\\\
In paper[3], They discuss and implement various RPL attacks in cooja simulator. They also present various solution present in the literature and give a new solution as Lightweight heartbeat protocol for RPL network. This new protocol is based on the successful transmission of ICMPv6 messages  from  the root node to nodes and vice versa. This new protocol gives less overhead but will work with its full potential if IPsec with ESP is used.\\\\
In paper[4], They describe the blackhole attack and present a solution for this in a 2-step process. In the first step, local decision is made by a node by observing the behaviour of neighbouring nodes. If any node is found to be suspicious, then the final decision whether this node is blackhole node or not taken by the root node, this process is called as global verification process. This solution is very effective but in this, every node observes the behaviour of its neighbours which increases the memory overload in these constrained devices and final decision is made by the root node which is a single point of failure(if compromised).\\\\
In paper[5], They describe the wormhole attack and present a solution for this as markle tree authentication. This solution is effective but increases the network complexity and control overhead due to the hashing and encryption techniques. In paper[6], They
 describe sinkhole attacks and give a solution which is the combination of two techniques as: Rank authentication technique(one way hash chaining) and parent fail-over technique(end to end acknowledgement). This solution is very promising but sybil attack is not managed by the parent fail-over technique and rank authentication is secure untill a random number given to the root node is not hacked by the attacker.\\\\
In paper[7], Authors presented SVELTE which is Real-time intrusion detection system for the
Internet of Things. This IDS is extensible and also uses the feature of firewall. But at a time implemented only few number of attacks detection, rest are still waiting to implement. But what if attackers hacked the security features of SVELTE? 
In paper[8], A hybrid routing protocol which is the combination of proactive and reactive approach is presented for wireless sensor networks with mobile sinks. This approach is very useful  for network life time and maintaining cost of DAGs.  \\\\
In paper[9], authors present a solution for rank and version number attacks as VeRA-version number and rank authentication in rpl. This
solution is cryptographically  secure  But their is some faulty results due to the missing correlation between  rank hash chain and version number hash chain. In paper[10], authors present solutions for mitigating topological attacks in RPL such as Vera++ and TRAIL. But both technology uses cryptography which makes these solutions highly complex and Distributed attackers communicated out-of-band channel cannot be detected in this model. \\\\
In paper[11], authors present a solution as SecTrust-RPL for mitigating the rank and sybil attacks. They use trust(direct or recommended) as a defence mechanism. This solution is very powerful but they assume the nodes are stationary and use location as a defence metric. They only simulate the direct trust in this model not the indirect one. They also assume only 10\% attacking nodes from the total nodes and  what if trust maintenance database is hacked by the attacker?\\\\
As shown from the above discussion, we conclude that every solution
has some limitation. This is the base of my motivation
for doing this research because there is no standard framework or
method to providing security in RPL network. Various
security modes such as unsecured, preinstalled, authenticated
are theoretically presented in standard RPL protocol.
But, the real time IoT products(produced by cisco and many
more) and simulation tools (cooja, RIOT and many more)
are still support only unsecured mode. Security modes are
still remains for practical implementation due to the constrained nature of Iot devices.

\section{Proposed Framework System}
We propose an ultimate approach (structure of proposed framework is shown in figure 3) for removing all internal and external RPL attacks. For removing internal attacks, we use a IDS or trust based approach and for mitigating external attacks, we have use some threshold values or rules in mini-firewall. We use intrusion detection system(combination of signature and anomaly based) at multiple locations (global and local) for analyzing the behaviour of nodes. Global IDS must be placed at 6LBR (sink node of rpl network) node and we take multiple fixed local sink node (approximately 10-15 \% of total nodes) along with local IDS, finally local sink nodes attach to the global sink node.\\ 
\begin{figure}[h]
\centering
\includegraphics[width=9cm,height=8.5cm]{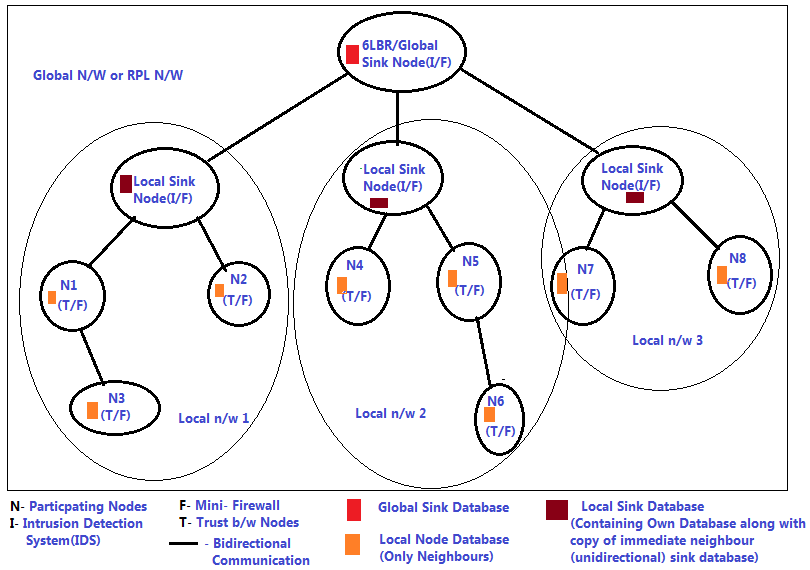}
    \caption{Proposed Framework}
    \label{fig:my_label}
\end{figure}

Iot devices or nodes are either mobile or stationary attach to the closest (closeness depends on the objective functions used as hop count, latency and energy minimization) local sink node. We also include the trust parameter(successful messages exchange ratio) between local nodes for better security. The final decision whether the node is attacker or not depends on mainly three  things as: trust between the neighbouring nodes, local decision by multiple sink nodes and global decision by root node. Mini-firewall is placed on every node as well as on every sink node and contains the list of authenticated nodes along with appropriate parameters such as ip address and rank for mitigating the external attacks.\\
\\In figure 3, every node including sink node maintain a database. For normal nodes, database contains the trust values and list of authenticated nodes (firewall rules) of each of their neighbour nodes(database is write protected or secure by the private key). Trust value is defined by successful packets exchange ratio such as more than 70 \% ratio is treated as best path, less than 40\% is treated as worst and in between them we check exactly three times(threshold value) for improving result or path. In each of local sink node, their are various tables or list with different functionalists. First table contain all of signature IDS rule, firewall rules and threshold values. Second thing we maintains a singly link list of blocks in which each block contains the transactions or packets of their local network according to particular time interval and block size which is useful for tracking of real time data(anomaly IDS).\\\\ For maintain immutability, each block contains the hash of the previous block same as blockchain. These block hashes and first table calculated hash are saved in second table which is write protected or secure by the private key. Both of the table update periodically whenever any new hash or rules are occurred correspondingly. For backup of data, we save the each of local sink link list to the immediate neighbour(unidirectional) local sink along with their hashes(update instantly). For global sink node, First table is same as the local sink node. Secondly, we maintain the combination of all link lists(update instantly) of each of local sink node for global view of the RPL network. Second table contains hash of first table along with hashes of all link lists for each of local network and rest is same as local sink node. Due to the resource constrained nature of devices, after some particular time interval we remove particular number of blocks from starting of link list along with their hashes in both local as well as global sink nodes while maintaining the summary of that records in the first table as adding some rules.\\\\ From the above we also say that, our proposed approach also follows the blockchain features such as immutability, decentralization, distributed and shared ledger. The more explanatory database structure is shown in figure 4:
\begin{figure}[h]
\centering
\includegraphics[width=9cm,height=9cm]{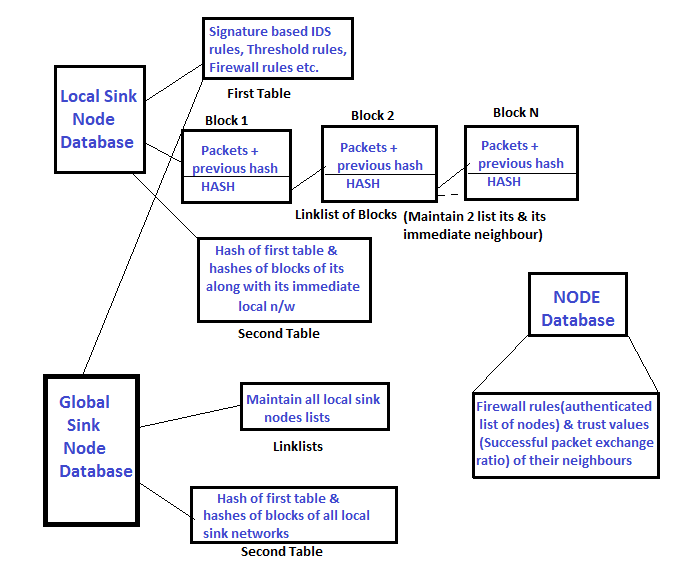}
    \caption{Database Structure}
    \label{fig:my_label}
\end{figure}

\section{Observations \& Theoretical Results }
The above proposed approach is theoretically analyzed in
the smart environment for mitigating the RPL attacks.
Some of the important analysis results \& observations are as follows:
\begin{itemize}
    \item It Removes the single point of failure(root node). As
in the proposed decentralized approach, The final decision whether the node is attacker or not depends on mainly three  things as: trust between the neighbouring nodes, local decision by multiple sink nodes and global decision by root node.\\
    \item  Due to the multiple local sink, number of transmission is
reduced from entire network to only closest sink and every local sink node maintains DAG only upto limited nodes in the main RPL network. So, it reduces network load in terms of control message
overhead and increased throughput.\\
    \item Sink discovery by sensor nodes almost eliminate use of
old routes which results in re-transmission of packets.
Thus, reduces network overhead.\\
    \item For removing unauthenticated devices and unauthenticated control packets(external attacks), we use mini-firewall at every node(including sink node) which maintains a database of authenticated devices using some parameters as ip address, rank, version number and sequence number.\\ 
     \item For mitigating the attacks occur from unencrypted control frames or compromised nodes (internal attacks), we use IDS which check the attacker at three phases. At first phase, we take trust parameter between nodes as mention above for selecting the best path (best parent node) towards sink. In the second phase, each local sink node maintain the link list of blocks for tracking the history of their network. Finally the global sink node maintain the history of DODAG (all local sink node) for mitigating any attack. The sink nodes also uses signature rules(as rank of parent node should be less than child node and many more rules) for easily detecting of known attacks.\\  
      \item  The link list of blocks is used for unknown attacks. For maintaining immutability of data, hash of previous block is contained in the current block same as blockchain by which we can detect any changes in that blocks at very less complexity. If we detect any changes, then from that block to the current block in that link list is replaced by the backup data(stored in the immediate neighbour sink node).\\
       \item Signature based rules, threshold values and firewall rules are periodically updated by which we take less time for detecting already happened attacks previously. \\
        \item Due to the resource limitations of sink nodes, after a particular time interval we remove certain number of blocks along with their hashes(in second table) from starting point of list while maintaining or adding that blocks important features as rules in the first table.\\
        \item All of the attacks are mitigated by this proposed approach by
seeing behavior of a node up
to joining of nodes into this network. After recognizing
of malicious node, it can be easily removed from
the network. Thus reduces DAG inconsistency and improve network performance.\\
         \item We assume all sink nodes including 6LBR are stationary. So, that we can check their geographical location for more better protection.\\
\item Overall, we use semi supervised IDS, Blockchain features, Trust parameters and mini-Firewall for protecting RPL network from various attacks.

\end{itemize}

\section{Conclusion \& Future work}
Finally, we say that RPL network(one of the most used routing protocol in IoT)is prone to various kind of attacks as rank, version modification attack and sybil attack etc. These attacks are very dangerous for the network resources and performance. These attacks are occur due to  problem of unauthenticated or unencrypted  control frames, centralized root controller, compromised  or unauthenticated devices and many more ways. There is no standard protection framework developed yet in the survey. So, we develop an ultimate framework for mitigating RPL attacks in smart environment. In this approach we use multiple sinks, Trust, semi supervised IDS and firewall mechanisms from protection to these attacks. We also maintain blockchain features for more better security. Then, we theoretically analyzed our approach which is very effective and reliable for mitigating these attacks. It also reduces network overhead along with increases the performance and throughput of the network due to the multiple sinks. These node's information are never compromised by the attacker due to the IDS, Firewall techniques and distributed blockchain features.\\

 This is our ongoing research work. In the future work, Firstly we will do simulation or real time implementation of this approach using network simulator (NS3) for RPL Secure network along with Ethereum blockchain.  After that we will focus on more attacks  such as Zero day attacks in RPL protocol and other layers attacks in IoT stack. We also want to reduce the complexity of our approach as low as possible through new ideas such as how to remove blocks of data at more effective way so that it can not loss any important information. Lastly we must ensure, our approach is backward compatible to original protocol.

\end{document}